\documentclass[conference]{IEEEtran}
\IEEEoverridecommandlockouts
\usepackage{cite}
\usepackage{amsmath,amssymb,amsfonts}
\usepackage{algorithmic}
\usepackage{graphicx}
\usepackage{textcomp}
\usepackage{xcolor}
\usepackage{pgfplots}
\pgfplotsset{compat=1.18}
\usepackage{pgfplotstable}

\def\BibTeX{{\rm B\kern-.05em{\sc i\kern-.025em b}\kern-.08em
    T\kern-.1667em\lower.7ex\hbox{E}\kern-.125emX}}

\begin{document}

\title{FedGenSC: Federated Generative Semantic Communication
with Channel-Aware Adaptation
\thanks{This work is supported by the American University of Beirut University Research Board (URB) and Vertically Integrated Projects (VIP) program. \\
© 2026 IEEE.Personal use of this material is permitted. Permission from IEEE must be obtained for all other uses, in any current or future media, including reprinting/republishing this material for advertising or promotional purposes, creating new collective works, for resale or redistribution to servers or lists, or reuse of any copyrighted component of this work in other works.}}

\author{Rita~Abou~Fares$^{\dagger}$,
        Razan~Al~Kakoun$^{\dagger}$,
        Maher~Nouiehed$^{*}$,
        Hadi~Sarieddeen$^{\dagger}$ \\
\small $^{\dagger}$Department of Electrical and Computer
Engineering, American University of Beirut, Beirut, Lebanon \\
$^{*}$Department of Industrial Engineering, American University
of Beirut, Beirut, Lebanon \\
\small rja62@mail.aub.edu,~rra65@mail.aub.edu,~%
maher.nouiehed@aub.edu.lb,~hadi.sarieddeen@aub.edu.lb}

\maketitle

\begin{abstract}

Integrating generative adversarial networks (GANs) into federated semantic communication (SemCom) is a natural progression, as generative priors can recover semantic fidelity under channel distortion that discriminative decoders cannot. However, naive GAN federation introduces three failure modes that prior work has, to the best of our knowledge, neither identified nor resolved: discriminator aggregation instability under non-independent and identically distributed (non-IID) data, semantic drift caused by divergent local embedding spaces, and channel-agnostic generation that cannot adapt to heterogeneous link conditions. We propose federated generative semantic communication (FedGenSC), which mitigates all three by employing a global generator with
local-only discriminators, providing cross-client semantic information through a semantic prototype bank, and conditioning generation on the
instantaneous signal-to-noise ratio (SNR).  Experiments on the Europarl dataset over Rayleigh fading channels ($K{=}10$ clients, Dirichlet $\alpha{=}0.5$) show that FedGenSC under non-IID data outperforms the FedDeepSC baseline across the tested SNR range, achieving up to a $58.2\%$ relative improvement in bilingual evaluation understudy (BLEU)-1 at 18\,dB. Ablation studies confirm the independent contribution of each component.

\end{abstract}


\section{Introduction}

The development of sixth-generation (6G) wireless networks has motivated communication paradigms that go beyond classical Shannon-theoretic bit-centric communication~\cite{b1}. Semantic communication (SemCom) has emerged as a promising approach that prioritizes the transmission of task-relevant information rather than exact bit sequences~\cite{b2}. When combined with joint source-channel coding (JSCC), SemCom can improve bandwidth efficiency and robustness under adverse channel conditions~\cite{b3}.

The DeepSC framework \cite{b3} established Transformer-based
semantic coding as a foundation for text-oriented SemCom, jointly optimizing
source and channel coding to preserve semantic information over noisy channels.
Recent work has further investigated DeepSC under realistic channel
conditions and semantic-aware adaptation to channel quality
\cite{IsmailTHzSemCom,IsmailSemanticSubBand}. However, envisioned 6G
applications are expected to involve distributed, privacy-sensitive, and
heterogeneous data, which can make centralized training impractical. Federated learning (FL)~\cite{b4} enables collaborative model optimization without exchanging raw data and has recently gained attention in SemCom~\cite{b5,b6,b7}. Recent studies have explored dynamic aggregation weighting for time-varying channels~\cite{b7}, loss-weighted partial updates~\cite{b8}, Byzantine-resilient aggregation~\cite{b5}, and trustworthy federated SemCom for Metaverse applications~\cite{b6}.

Despite these advances, existing federated SemCom frameworks rely
on discriminative decoders and standard aggregation strategies.
Incorporating GANs~\cite{b10} is a natural next step, as
generative priors can recover semantic fidelity under channel
distortion that discriminative decoders may not fully mitigate~\cite{b9}.
In centralized settings, GANs have been shown to suppress channel
distortion without requiring full channel state information
(CSI)~\cite{b15}. Federated GAN training has also been explored in
non-SemCom contexts~\cite{b11}, without addressing the semantic
and channel-specific deployment challenges.

However, extending GAN-based SemCom approaches to the federated
setting introduces three fundamental challenges:
(\textbf{P1}) discriminator aggregation instability under non-independent
and identically distributed (non-IID) data can lead to generator collapse;
(\textbf{P2}) semantic drift, whereby divergent local embedding spaces
can degrade the aggregated generator's performance on local clients;
and (\textbf{P3}) channel-agnostic generation, whereby a generator
that is unaware of the instantaneous channel state cannot pre-adapt
its output to heterogeneous link conditions. To the best of our knowledge, these three problems have not been jointly characterized and addressed in prior federated SemCom literature, forming the core contribution of this work.

A natural solution is to introduce a semantic pre-coding component, a
module that rectifies distortions arising from channel noise before
transmission. However, such a component cannot be naively deployed at
local clients in a federated setting: under non-IID data and
heterogeneous channels, each client would learn a locally biased
corrector, and aggregating these correctors could be unstable.
We propose a GAN-based structure in which only the generator is
federated, acting as a shared semantic pre-equalizer, while discriminators
remain local. Classical signal-domain precoders~\cite{b16} typically rely on full
channel state information (CSI), while neural network
precoders~\cite{b17} learn precoding mappings from CSI. In contrast, the proposed
federated generative semantic communication (FedGenSC) pre-equalizes
in the \emph{semantic embedding space} using only a scalar
signal-to-noise ratio (SNR) estimate, a
distinction that makes it well suited to federated,
privacy-constrained deployments.

To address P1--P3 jointly, the proposed FedGenSC builds on three
principled design elements: generator-only federation with
local-only discriminators, a semantic prototype bank for
cross-client embedding grounding, and channel-state-conditioned
generation. Experiments on the Europarl dataset under Rayleigh
fading show that FedGenSC addresses the three identified failure
modes of GAN federation, with the contribution of each design
element independently evaluated through controlled ablation studies.
\section{Problem Formulation}
\label{sec:problem_formulation}

\subsection{System Model}

Consider a federated SemCom system with one central server and
$K$ clients indexed by $k \!\in\! \{1,\ldots,K\}$. Each client $k$
holds a local text dataset
$\mathcal{D}_k \!=\! \{\mathbf{s}_i^{(k)}\}_{i=1}^{N_k}$, where each
sentence $\mathbf{s}_i^{(k)}$ is drawn from a local distribution
$\mathcal{P}_k$. In the non-IID setting,
$\mathcal{P}_i \!\neq\! \mathcal{P}_j$ for $i \!\neq\! j$~\cite{b12},
meaning that clients hold data drawn from different local distributions.
Each client operates a local instance of 
DeepSC~\cite{b3}, a Transformer-based semantic codec. The semantic
encoder $f_{\boldsymbol{\alpha}}$ maps an input sentence
$\mathbf{s}$, tokenized to length $T$, to a sequence of semantic
embeddings
$\mathbf{E} \!=\! f_{\boldsymbol{\alpha}}(\mathbf{s}) \!\in\!
\mathbb{R}^{T \times d}$, where $d{=}128$ is the model dimension
and $\mathbf{e}_t \!\in\! \mathbb{R}^d$ is the semantic representation
of the $t$-th token. The channel encoder $f_{\boldsymbol{\beta}}$
compresses these embeddings from $d{=}128$ to $16$ dimensions per
token for transmission. At the receiver, the combined channel and
semantic decoder $g_{\boldsymbol{\delta}}$ reconstructs the
original sentence $\hat{\mathbf{s}}$ from the received signal.

Each client $k$ experiences an independent Rayleigh fading channel
with instantaneous SNR $\gamma_k$, computed from the noise
variance $\sigma_k^2$ as
\begin{equation}
    \gamma_k^{\mathrm{dB}} =
    10\log_{10}\!\left(\frac{1}{2\sigma_k^2}\right),
    \label{eq:snr}
\end{equation}
where the transmitted signal is normalized to unit average power,
$\mathbb{E}[\|\mathbf{x}_k\|^2]=1$, and the factor of $2$ accounts
for $\sigma_k^2$ denoting the noise variance per real (in-phase or
quadrature) dimension in Eq.~\eqref{eq:channel}, so that the total
per-dimension complex noise power is $2\sigma_k^2$. The received
signal is
\begin{equation}
    \mathbf{y}_k = \mathbf{h}_k \odot \mathbf{x}_k + \mathbf{n}_k,
    \label{eq:channel}
\end{equation}
where $\mathbf{x}_k$ is the power-normalized transmitted signal,
$\odot$ denotes element-wise multiplication,
$\mathbf{h}_k \sim \mathcal{CN}(\mathbf{0},\mathbf{I})$ is the
Rayleigh fading vector, and
$\mathbf{n}_k \sim \mathcal{CN}(\mathbf{0},2\sigma_k^2\mathbf{I})$
is additive white Gaussian noise (AWGN).

\subsection{Federated Training Objective}

The global objective is to find the shared parameters
$\boldsymbol{\omega} = \{\boldsymbol{\alpha}, \boldsymbol{\beta},
\boldsymbol{\delta}, \boldsymbol{\theta}\}$, where
$\boldsymbol{\alpha}$, $\boldsymbol{\beta}$, $\boldsymbol{\delta}$,
and $\boldsymbol{\theta}$ denote the semantic encoder, channel
encoder, decoder, and generator parameters, respectively, by
minimizing the expected semantic distortion across all clients:
\begin{equation}
    \min_{\boldsymbol{\omega}} \; \frac{1}{K}\sum_{k=1}^{K}
    \mathbb{E}_{\mathbf{s}\sim\mathcal{P}_k}\!\left[
    L_{\mathrm{sem}}\!\left(\mathbf{s},\,\hat{\mathbf{s}}\right)
    \right],
    \label{eq:global_obj}
\end{equation}
subject to the constraint that no raw data $\mathcal{D}_k$ are
shared~\cite{b4,b14}. Here,
$L_{\mathrm{sem}}(\mathbf{s},\hat{\mathbf{s}})
\!=\! -\!\sum_t s_t \log \hat{s}_t$ is the token-level cross-entropy
loss. BLEU-1 is used in Section~\ref{sec:Emperical_results} to
evaluate the reconstruction quality obtained by minimizing this loss.
The privacy constraint renders centralized DeepSC an unsuitable
comparison as it requires pooling all $\mathcal{D}_k$ on a
single server.

\subsection{Challenges of Integrating GANs into Federated SemCom}
\label{subsec:challenges}

\subsubsection{P1 --- Discriminator Aggregation Instability}
Augmenting DeepSC with a generator
$G_{\boldsymbol{\theta}}$ and discriminator $D_{\boldsymbol{\phi}}$,
and aggregating both via federated averaging (FedAvg)~\cite{b4}, can lead to instability.
Under non-IID data, local discriminators
$D_{\boldsymbol{\phi}_k}$ develop incompatible decision
boundaries. Naively averaging their parameters,
$\bar{\boldsymbol{\phi}} = \frac{1}{K}\sum_k \boldsymbol{\phi}_k$,
produces a global discriminator $D_{\bar{\boldsymbol{\phi}}}$ whose
decision boundary need not correspond to any client's local minimax
solution,
\begin{equation}
    \min_{G}\max_{D_k}\;L_{\mathrm{adv}}(G,D_k).
    \label{eq:minimax}
\end{equation}
Empirically, discriminator aggregation degrades BLEU-1 by $61.8\%$
at $6$\,dB and falls \emph{below} the no-GAN baseline at
mid-to-high SNR.

\subsubsection{P2 --- Semantic Drift}
Under non-IID distributions, local embedding spaces
$\{f_{\boldsymbol{\alpha}}(\mathcal{D}_k)\}_{k=1}^K$ can diverge
as each client's encoder is fine-tuned on a different data
distribution, so the same word or phrase can map to different
regions of $\mathbb{R}^d$ across clients. The aggregated generator
may consequently produce embeddings that deviate from the underlying
client representations, a phenomenon we denote as \emph{semantic
drift}. We show empirically that addressing this problem yields a
$62.8\%$ improvement in BLEU-1 at $6$\,dB
(Section~\ref{sec:Emperical_results}).

\subsubsection{P3 --- Channel-Agnostic Generation}
Clients operate under heterogeneous instantaneous SNRs $\gamma_k$.
Let $\bar{\mathbf{e}} \in \mathbb{R}^d$ denote the mean-pooled
sentence embedding (check Eq.~\eqref{eq:mean_pool}). An idealized
channel-conditioned generator objective is
\begin{equation}
    G^*(\bar{\mathbf{e}},\gamma) =
    \underset{G}{\arg\min}\;
    \mathbb{E}_{\mathbf{h},\mathbf{n}}\!\left[
    L_{\mathrm{sem}}\!\left(\mathbf{s},\,
    g_{\boldsymbol{\delta}}\!\left(
    \mathbf{y}(\gamma,G(\bar{\mathbf{e}},\gamma))
    \right)\right)\right],
    \label{eq:optimal_gen}
\end{equation}
where $\mathbf{y}$ is the received signal (Eq.~\eqref{eq:channel}),
which also depends on the channel encoder; this dependence is
suppressed here for notational simplicity.
A channel-agnostic generator $G(\bar{\mathbf{e}})$ neglects the
dependence on $\gamma$. As a result, it can under-correct in low-SNR
regimes and over-correct in high-SNR conditions, thereby distorting
otherwise clean embeddings.
In Section~\ref{sec:Emperical_results}, we quantify the effect of
addressing each of P1--P3 through controlled ablation, showing that
SNR conditioning reduces BLEU-1 degradation by up to $49.4\%$ at
$6$\,dB.

\section{Proposed Framework: FedGenSC}

\subsection{Overview and Design Principles}

FedGenSC incorporates a conditional GAN that operates
directly in the semantic embedding space. The generator is introduced
after the semantic encoder: it takes the mean-pooled sentence embedding
$\bar{\mathbf{e}}$ as input, along with the instantaneous SNR
$\gamma_k^{\mathrm{dB}}$ and a semantic prototype vector $\mathbf{p}$,
and produces a residual correction, a semantic correction vector, added to the original embedding prior to channel encoding.

The discriminator operates in the semantic embedding space, prior to
channel encoding. It receives as input both the original mean-pooled
embedding $\bar{\mathbf{e}}$ and the generator-corrected embedding
$\tilde{\mathbf{e}}$, and is trained to distinguish between them. This
design enables the generator to learn corrections that produce
embeddings statistically indistinguishable from clean embeddings,
which in turn improves robustness once these corrected embeddings are
transmitted over the noisy channel.
Operating in the semantic embedding space, rather than the raw
signal space, reduces the complexity of learning meaningful
corrections, since the embedding space is lower-dimensional than
raw signals with high-dimensional noise patterns less directly
aligned with semantic structure.

\subsection{System Architecture}
\label{subsec:arch}

The FedGenSC pipeline interposes a conditional GAN between the
DeepSC semantic and channel encoders. Throughout,
$B$ denotes the batch size, $T$ denotes the sequence length, and
$d{=}128$ denotes the model dimension. The Transformer encoder
$f_{\boldsymbol{\alpha}}$ maps the tokenized input to
$\mathbf{E} \in \mathbb{R}^{B \times T \times d}$, whose $t$-th
row (per batch element) is the token embedding
$\mathbf{e}_t \in \mathbb{R}^d$ introduced in
Section~\ref{sec:problem_formulation}.
A sentence-level representation is obtained by mean pooling:
\begin{equation}
    \bar{\mathbf{e}} = \frac{1}{T}\sum_{t=1}^{T}\mathbf{e}_t
    \in \mathbb{R}^{B \times d}.
    \label{eq:mean_pool}
\end{equation}
The generator $G_{\boldsymbol{\theta}}$ is a three-layer multilayer
perceptron (MLP)
($2d{+}1 \to 512 \to 256 \to d$, $\approx$296K parameters,
${<}5\%$ of DeepSC). It takes the concatenated input
$[\bar{\mathbf{e}},\,\gamma_k^{\mathrm{dB}},\,\mathbf{p}]
\in \mathbb{R}^{2d+1}$ and produces a residual correction
$\hat{\mathbf{e}} \in \mathbb{R}^{B \times d}$ that is added to
the encoder output:
\begin{equation}
    \tilde{\mathbf{E}} = \mathbf{E} +
    \hat{\mathbf{e}}\cdot\mathbf{1}^{\top},
    \quad \tilde{\mathbf{E}} \in \mathbb{R}^{B \times T \times d},
    \label{eq:residual}
\end{equation}
where $\mathbf{1} \in \mathbb{R}^T$ is an all-ones vector and
$(\cdot)^{\top}$ denotes vector transpose, so that
$\hat{\mathbf{e}}\cdot\mathbf{1}^{\top} \in \mathbb{R}^{B\times T\times d}$
broadcasts the correction across all $T$ token positions.

The discriminator $D_{\boldsymbol{\phi}_k}$ is a two-layer MLP with
spectral normalization and a sigmoid activation. It takes as input
$[\bar{\mathbf{e}},\,\tilde{\mathbf{e}}] \in \mathbb{R}^{2d}$, where
$\tilde{\mathbf{e}} = \bar{\mathbf{e}} + \hat{\mathbf{e}}$ is the
generator-corrected embedding, and returns a probability in $(0,1)$.
The discriminator is instantiated locally in each round and discarded
after local training, directly addressing~(P1). The corrected
$\tilde{\mathbf{E}}$ is then passed to the channel encoder and
transmitted over the Rayleigh fading channel
(Eq.~\eqref{eq:channel}) for decoding by
$g_{\boldsymbol{\delta}}$.
\smallskip

\subsection{Semantic Prototype Bank}
\label{subsec:proto}

To address (P2), we introduce a lightweight mechanism that aligns
heterogeneous client semantic distributions without sharing raw data
or gradients. Each client $k$ constructs a compact summary of its
local semantic space by performing $k$-means clustering over the
sentence embeddings collected during the local training round:
\begin{equation}
\mathcal{B}_k =
\mathrm{KMeans}\!\left(\{\bar{\mathbf{e}}_i^{(k)}\},\,P\right),
\label{eq:kmeans}
\end{equation}
where $P$ is the number of centroids and
$\bar{\mathbf{e}}_i^{(k)} \in \mathbb{R}^d$ denotes the
$i$-th sentence embedding at client $k$. Only these $P$ centroid
vectors are transmitted to the server, reducing communication
overhead while avoiding raw data sharing.

The server combines all local summaries into a global prototype bank
$\mathcal{B} = \bigcup_{k=1}^K \mathcal{B}_k$, which is
redistributed to all clients at the beginning of each round.

For each input embedding $\bar{\mathbf{e}}$, a prototype vector is
selected as the centroid in the global bank with the highest cosine
similarity:
\begin{equation}
\mathbf{p} =
\underset{\mathbf{c}\in\mathcal{B}}{\arg\max}\;
\frac{\bar{\mathbf{e}}\cdot\mathbf{c}}
{\|\bar{\mathbf{e}}\|\,\|\mathbf{c}\|},
\quad \mathbf{p} \in \mathbb{R}^{B \times d},
\label{eq:prototype}
\end{equation}
where the nearest-centroid search is performed independently for
each of the $B$ embeddings in a batch, yielding one prototype vector
per sample.
This prototype serves as a semantic anchor, providing global
context about the position of the embedding relative to the union
of client distributions: it helps counteract misaligned embedding
geometries that contribute to semantic drift, consistent with prior
prototype-based approaches~\cite{b18,b20}, while conditioning the
generator toward a globally consistent semantic space
to improve robustness and generalization across clients.
This mechanism introduces limited communication overhead, as
each client transmits only $P \times d$ floating-point values per
round (e.g., $2{,}560$ values for $P=20$ and $d=128$), linear in
$P$ and $d$.

\subsection{Two-Timescale Local Training Protocol}

In each communication round, client $k$ executes two sequential
phases. In the first phase, DeepSC is fine-tuned on the local dataset
$\mathcal{D}_k$ for $E{=}3$ epochs using the semantic loss
$L_{\mathrm{sem}}$. During this phase, the generator
$G_{\boldsymbol{\theta}}$ acts as a residual adapter, producing a
correction that is added to the encoder output prior to channel
encoding. In the second phase, the encoder parameters are fixed so
that GAN training operates on a stable embedding space. A fresh
discriminator $D_{\boldsymbol{\phi}_k}$ is instantiated, while the
generator is initialized from the global model
$G_{\bar{\boldsymbol{\theta}}}$. A two-timescale update scheme is
used, with $n_D{=}5$ discriminator steps per generator step.
Let
\begin{equation}
    \hat{\mathbf e}
    =
    G_{\boldsymbol{\theta}_k}
    \left(
    \bar{\mathbf e},\gamma_k^{\mathrm{dB}},\mathbf p
    \right),
    \qquad
    \tilde{\mathbf e}
    =
    \bar{\mathbf e}
    +
    \hat{\mathbf e},
\end{equation}
denote the generator's residual correction and the resulting
corrected embedding, respectively.  The discriminator receives
a reference clean embedding $\bar{\mathbf e}$ and a candidate
embedding and is trained to distinguish real semantic pairs from
generated ones. The real pair is
$(\bar{\mathbf e},\bar{\mathbf e})$, while the generated pair is
$(\bar{\mathbf e},\tilde{\mathbf e})$. The discriminator loss is
\begin{align}
    L_D
    &=
    -\mathbb{E}_{\bar{\mathbf e}}
    \left[
    \log
    D_{\boldsymbol{\phi}_k}
    \left(
    \bar{\mathbf e},\bar{\mathbf e}
    \right)
    \right]
    \nonumber\\
    &\quad
    -
    \mathbb{E}_{\bar{\mathbf e},\gamma_k^{\mathrm{dB}},\mathbf p}
    \left[
    \log
    \left(
    1-
    D_{\boldsymbol{\phi}_k}
    \left(
    \bar{\mathbf e},
    \tilde{\mathbf e}
    \right)
    \right)
    \right].
    \label{eq:disc_loss}
\end{align}

The generator is trained to produce corrections that remain close to
the original embedding while making the corrected embeddings
indistinguishable from real semantic embeddings. Its loss is
\begin{equation}
    L_G =
    \underbrace{
    \mathbb{E}
    \left[
    \|\tilde{\mathbf e} - \bar{\mathbf e}\|_2^2
    \right]
    }_{\text{residual regularization}}
    -
    \lambda_1
    \underbrace{
    \mathbb{E}
    \left[
    \log
    D_{\boldsymbol{\phi}_k}
    \left(
    \bar{\mathbf e},\tilde{\mathbf e}
    \right)
    \right]
    }_{\text{adversarial alignment}}.
    \label{eq:gen_loss}
\end{equation}
The residual regularization term prevents excessive deviation from
the original embedding, while the adversarial term encourages the
corrected embedding to lie on the manifold of valid sentence
representations.

\vspace{0.5em}
\noindent
\textbf{Prototype Alignment (Global Consistency).}
To mitigate semantic drift across clients, prototype alignment is
handled explicitly through the shared prototype bank. Let
\begin{equation}
    \bar{\mathbf p}_k
    =
    \frac{1}{P}
    \sum_{j=1}^P
    \mathbf c_j^{(k)}, \qquad
    \bar{\mathbf p}
    =
    \frac{1}{K}
    \sum_{k=1}^K
    \bar{\mathbf p}_k,
    \label{eq:mean_proto}
\end{equation}
denote client $k$'s local mean prototype and the global mean
prototype, respectively; these are used below to weight the
generator aggregation (Section~\ref{subsec:aggregation}).
Local prototype banks $\mathcal{B}_k$ are periodically aggregated at
the server, providing common prototype information across clients.

\vspace{0.5em}
\noindent
After local training, client $k$ transmits to the server
(i) the updated DeepSC parameters,
(ii) the updated generator parameters $\boldsymbol{\theta}_k$,
(iii) the representative training SNR $\gamma_k$, and
(iv) the local prototype bank $\mathcal{B}_k$.
The discriminator is discarded and never transmitted.

\subsection{Federated Generator Aggregation}
\label{subsec:aggregation}
The server aggregates the generator parameters
$\{\boldsymbol{\theta}_k\}_{k=1}^K$, while the local discriminator
parameters are neither transmitted nor aggregated. This directly
targets (P1) by decoupling discriminator training across clients.

The global generator is computed as a weighted average:
\begin{equation}
    \bar{\boldsymbol{\theta}} =
    \sum_{k=1}^{K} w_k \boldsymbol{\theta}_k,
\end{equation}
where the weights $\{w_k\}$ reflect the reliability of each
client. In heterogeneous settings, naive averaging can be
suboptimal because clients may operate under different SNR regimes,
learn generators of varying quality, or exhibit misaligned semantic
structures.

To account for these differences, each weight is constructed from
three complementary quality signals:
\begin{equation}
    s_k^{\mathrm{SNR}} =
    \frac{1}{1 + |\gamma_k - \bar{\gamma}|}, \quad
    s_k^{\mathrm{gen}} =
    \frac{1}{1 + \mathcal{L}_k^{\mathrm{recon}}},
\end{equation}
\begin{equation}
    s_k^{\mathrm{proto}} =
    \max\!\left(0,\,
    \frac{
    \bar{\mathbf p}_k \cdot
    \bar{\mathbf p}
    }{
    \|\bar{\mathbf p}_k\|\,
    \|\bar{\mathbf p}\|
    }\right),
\end{equation}
where $\bar{\gamma} = \frac{1}{K}\sum_k \gamma_k$, $\bar{\mathbf p}_k$
and $\bar{\mathbf p}$ are the local and global mean prototypes defined
in Eq.~\eqref{eq:mean_proto}, and
$\mathcal{L}_k^{\mathrm{recon}} =
\mathbb{E}\!\left[\|\tilde{\mathbf e}-\bar{\mathbf e}\|_2^2\right]$
is the residual regularization term of Eq.~\eqref{eq:gen_loss},
evaluated on client $k$'s local data.

The final weights are normalized:
\begin{equation}
    w_k =
    \frac{
    s_k^{\mathrm{SNR}} \cdot
    s_k^{\mathrm{gen}} \cdot
    s_k^{\mathrm{proto}}
    }{
    \sum_{j=1}^K
    s_j^{\mathrm{SNR}} \cdot
    s_j^{\mathrm{gen}} \cdot
    s_j^{\mathrm{proto}}
    }.
\end{equation}
This weighting favors clients whose channel conditions are
representative ($s_k^{\mathrm{SNR}}$), whose generators produce
accurate corrections ($s_k^{\mathrm{gen}}$), and whose semantic
structures align with the global space ($s_k^{\mathrm{proto}}$).
By combining these signals, the aggregation is designed to reduce
the influence of outlying clients and limit overfitting to a single
SNR regime or local data distribution.

Table~\ref{tab:solutions} summarizes the proposed solutions to the
challenges introduced by data and channel heterogeneity.

\begin{table}[!t]
\renewcommand{\arraystretch}{1.3}
\caption{Problem--Solution Mapping in FedGenSC}
\label{tab:solutions}
\centering
\begin{tabular}{p{2.4cm}|p{5.4cm}}
\hline
\textbf{Problem} & \textbf{FedGenSC Solution} \\
\hline
P1: GAN instability &
Local-only discriminator; generator weights only are shared;
discriminator re-initialized each round. \\
\hline
P2: Semantic drift &
Prototype bank ($k$-means centroids shared, not raw data); prototype-similarity weight
$s_k^{\mathrm{proto}}$ in aggregation. \\
\hline
P3: Channel-agnostic generation &
Generator conditioned on $\gamma_k^{\mathrm{dB}}$ at every
forward pass; SNR-similarity weight $s_k^{\mathrm{SNR}}$
in aggregation. \\
\hline
\end{tabular}
\end{table}



\section{Empirical Results}
\label{sec:Emperical_results}

\subsection{Experimental Setup}

Experiments use the Europarl parallel corpus~\cite{b3}
($|\mathcal{V}|{=}22{,}234$, batch size $B{=}128$). The federated
system comprises $K{=}10$ clients with data partitioned via a
Dirichlet distribution with concentration $\alpha{=}0.5$, representing a moderately heterogeneous non-IID setting. Each client runs $E{=}3$ local epochs in Phase~1 and
at least $30$ GAN update steps in Phase~2, over 50 rounds. The prototype bank uses $P{=}20$ centroids per client. The channel model is
Rayleigh fading (Eq.~\eqref{eq:channel}) evaluated at SNR
$\in \{0,3,6,9,12,15,18\}$\,dB. Both backbone and GAN use the
Adam optimizer with $\eta{=}10^{-4}$, $\beta_1{=}0.9$,
$\beta_2{=}0.98$. Performance is measured by BLEU-1 (lexical
fidelity) and SBERT cosine similarity~\cite{b13} (semantic
fidelity).

The FedDeepSC baselines (IID and non-IID) are identical to
FedGenSC with the GAN module removed, so that observed performance
differences can be attributed to the proposed GAN components, though
part of the gain may also reflect the added model capacity of the
GAN module (see the note in Section~\ref{sec:Emperical_results}).
Centralized DeepSC is excluded as a baseline because
it requires pooling all client data on a single server,  which is
precluded by the no-raw-data-sharing constraint underlying
Eq.~\eqref{eq:global_obj}.

\subsection{Main Experiments}

Tables~\ref{tab:main_results_full} and~\ref{tab:ablation_full}
report BLEU-1 scores across the full SNR range under Rayleigh
fading.

\begin{table*}[t]
\centering
\begin{minipage}{0.48\textwidth}
    \centering
    \caption{BLEU-1 Scores Across Full SNR Range Under Rayleigh
    Fading. Non-IID: Dirichlet $\alpha{=}0.5$, $K{=}10$ clients.}
    \label{tab:main_results_full}
    \small
    \begin{tabular}{c|cc|cc|c}
    \hline
    \textbf{SNR} & \textbf{Base.} & \textbf{GAN} &
    \textbf{Base.} & \textbf{GAN} & \textbf{Naive} \\
    \textbf{(dB)} & \textbf{IID} & \textbf{IID} &
    \textbf{N-IID} & \textbf{N-IID} & \textbf{GAN} \\
    \hline
    0  & 0.1696 & 0.3590 & 0.1577 & 0.4265 & 0.1550 \\
    3  & 0.2138 & 0.4354 & 0.1921 & 0.5290 & 0.1794 \\
    6  & 0.2772 & 0.4940 & 0.2165 & 0.6028 & 0.2303 \\
    9  & 0.3554 & 0.5368 & 0.2469 & 0.6480 & 0.3012 \\
    12 & 0.4236 & 0.5631 & 0.3304 & 0.6715 & 0.3523 \\
    15 & 0.4743 & 0.5803 & 0.3872 & 0.6854 & 0.4028 \\
    18 & 0.5097 & 0.5920 & 0.4397 & 0.6956 & 0.4336 \\
    \hline
    \end{tabular}
\end{minipage}
\hfill
\begin{minipage}{0.48\textwidth}
    \centering
    \caption{Ablation Study: BLEU-1 Across Full SNR Range.
    All runs: Non-IID, $K{=}10$ clients.}
    \label{tab:ablation_full}
    \small
    \addtolength{\tabcolsep}{-2pt}
    \begin{tabular}{lcccc}
    \hline
    \textbf{System} &
    \textbf{0\,dB} & \textbf{6\,dB} &
    \textbf{12\,dB} & \textbf{18\,dB} \\
    \hline
    
    Full FedGenSC  & 0.4265 & 0.6028 & 0.6715 & 0.6956 \\
    Naive GAN Fed. & 0.1550 & 0.2303 & 0.3523 & 0.4336 \\
    w/o Prototype  & 0.1500 & 0.2241 & 0.4008 & 0.5502 \\
    w/o SNR Cond.  & 0.1609 & 0.3048 & 0.4909 & 0.6103 \\
    Baseline N-IID & 0.1577 & 0.2165 & 0.3304 & 0.4397 \\
    \hline
    \end{tabular}
    \vspace{4pt}
    \begin{flushleft}
    \scriptsize
    \textit{Drop vs.\ full at 6\,dB:
    Naive GAN $-61.8\%$;
    w/o Proto.\ $-62.8\%$;
    w/o SNR Cond.\ $-49.4\%$.}
    \end{flushleft}
\end{minipage}
\end{table*}

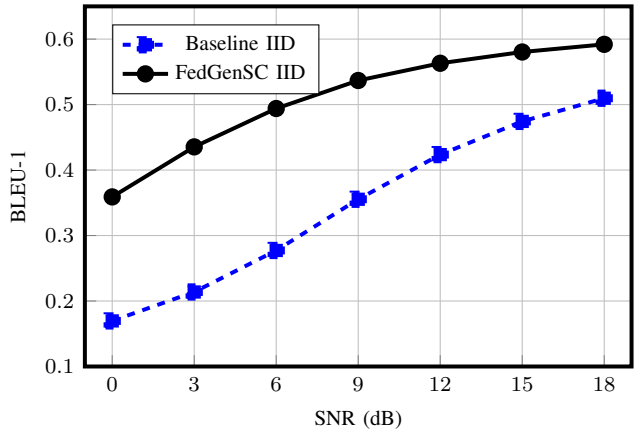
\begin{figure}[t]
\centering
\begin{tikzpicture}
\begin{axis}[
    width=\columnwidth,
    height=0.72\columnwidth,
    xlabel={SNR (dB)},
    ylabel={BLEU-1},
    xlabel style={font=\footnotesize},
    ylabel style={font=\footnotesize},
    tick label style={font=\footnotesize},
    legend style={font=\footnotesize,
        at={(0.05,0.95)}, anchor=north west,
        draw=black, line width=0.4pt},
    xmin=-1, xmax=19,
    ymin=0.10, ymax=0.65,
    xtick={0,3,6,9,12,15,18},
    ytick={0.1,0.2,0.3,0.4,0.5,0.6,0.7},
    grid=both,
    grid style={line width=0.3pt,draw=gray!30},
    major grid style={line width=0.4pt,draw=gray!50},
    tick pos=left, line width=1.5pt]
\addplot[color=blue,dashed,line width=1.5pt,mark=square*,
    mark size=2.5pt,mark options={fill=blue,draw=blue}]
    coordinates{(0,0.1696)(3,0.2138)(6,0.2772)(9,0.3554)
                (12,0.4236)(15,0.4743)(18,0.5097)};
\addlegendentry{Baseline IID}
\addplot[color=black,solid,line width=1.5pt,mark=*,
    mark size=2.5pt,mark options={fill=black,draw=black}]
    coordinates{(0,0.3590)(3,0.4354)(6,0.4940)(9,0.5368)
                (12,0.5631)(15,0.5803)(18,0.5920)};
\addlegendentry{FedGenSC IID}
\end{axis}
\end{tikzpicture}
\caption{BLEU-1 vs.\ SNR under IID partitioning. FedGenSC
consistently outperforms FedDeepSC across the full SNR range, with larger gains at lower SNR where channel noise is stronger.
Rayleigh fading, $K{=}10$ clients.}
\label{fig:iid_comparison}
\end{figure}

\begin{figure}[t]
\centering
\begin{tikzpicture}
\begin{axis}[
    width=\columnwidth,
    height=0.75\columnwidth,
    xlabel={SNR (dB)},
    ylabel={BLEU-1},
    xlabel style={font=\footnotesize},
    ylabel style={font=\footnotesize},
    tick label style={font=\footnotesize},
    legend style={font=\footnotesize,
        at={(0.05,0.95)}, anchor=north west,
        draw=black, line width=0.4pt},
    xmin=-1, xmax=19,
    ymin=0.10, ymax=0.72,
    xtick={0,3,6,9,12,15,18},
    ytick={0.1,0.2,0.3,0.4,0.5,0.6,0.7},
    grid=both,
    grid style={line width=0.3pt,draw=gray!30},
    major grid style={line width=0.4pt,draw=gray!50},
    tick pos=left, line width=1.5pt]
\addplot[color=blue,dashed,line width=1.5pt,mark=square*,
    mark size=2.5pt,mark options={fill=blue,draw=blue}]
    coordinates{(0,0.1577)(3,0.1921)(6,0.2165)(9,0.2469)
                (12,0.3304)(15,0.3872)(18,0.4397)};
\addlegendentry{Baseline Non-IID}
\addplot[color=black,solid,line width=1.5pt,mark=*,
    mark size=2.5pt,mark options={fill=black,draw=black}]
    coordinates{(0,0.4265)(3,0.5290)(6,0.6028)(9,0.6480)
                (12,0.6715)(15,0.6854)(18,0.6956)};
\addlegendentry{FedGenSC Non-IID}
\end{axis}
\end{tikzpicture}
\caption{BLEU-1 vs.\ SNR under non-IID partitioning (Dirichlet
$\alpha{=}0.5$, $K{=}10$ clients). FedGenSC exceeds the
non-IID baseline across the tested SNR range. Rayleigh fading.}
\label{fig:noniid_comparison}
\end{figure}
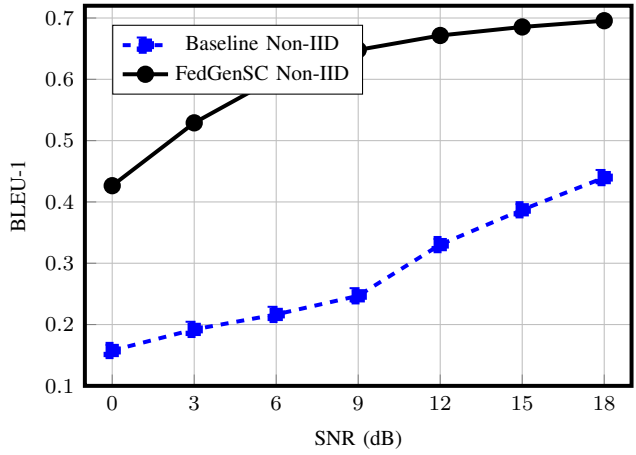

\paragraph{Non-IID setting}
Moving from IID to non-IID degrades the FedDeepSC baseline by
$21.9\%$ at 6\,dB, confirming that data heterogeneity is a
persistent challenge in federated SemCom. FedGenSC substantially
recovers this loss, exceeding both the non-IID and IID baselines
across the full SNR range and reaching a $58.2\%$ relative
improvement over the non-IID baseline at 18\,dB
(Fig.~\ref{fig:noniid_comparison}), reflecting that the generator
provides effective semantic corrections across the tested channel conditions. Part of this gain may also reflect the GAN module's added model
capacity rather than non-IID resolution alone. SBERT scores are
similarly higher in the non-IID setting (Table~\ref{tab:sbert_results});
without a non-IID SBERT baseline, we do not treat this as an
isolated prototype-alignment gain, though it is consistent with
stronger regularization under heterogeneous distributions.
\paragraph{IID setting}
Fig.~\ref{fig:iid_comparison} shows consistent improvement over the
IID baseline across all SNR values, with larger gains at low-to-mid
SNR where the GAN correction is most beneficial. This suggests that the prototype bank contribution is additive: in the non-IID case it provides semantic grounding that helps maintain gains across the full SNR range, while in the IID case the benefit diminishes at high SNR as smaller channel-induced distortions require less correction.

Table~\ref{tab:sbert_results} reports BLEU-1 and SBERT scores, showing consistent lexical and semantic gains across SNR; since SBERT measures embedding similarity beyond direct word overlap, these results indicate that the BLEU-1 gains are accompanied by improved semantic fidelity rather than merely lexical matching.

\begin{table}[t]
\centering
\caption{BLEU-1 and SBERT Similarity vs.\ SNR for FedGenSC.}
\label{tab:sbert_results}
\footnotesize
\begin{tabular}{c|cc|cc}
\hline
& \multicolumn{2}{c|}{\textbf{IID}}
& \multicolumn{2}{c}{\textbf{Non-IID} ($\alpha{=}0.5$)} \\
\textbf{SNR (dB)} &
\textbf{BLEU-1} & \textbf{SBERT} &
\textbf{BLEU-1} & \textbf{SBERT} \\
\hline
0  & 0.3590 & 0.4186 & 0.4265 & 0.4646 \\
3  & 0.4354 & 0.4630 & 0.5290 & 0.5272 \\
6  & 0.4940 & 0.4979 & 0.6028 & 0.5699 \\
9  & 0.5368 & 0.5222 & 0.6480 & 0.5972 \\
12 & 0.5631 & 0.5382 & 0.6715 & 0.6115 \\
15 & 0.5803 & 0.5482 & 0.6854 & 0.6193 \\
18 & 0.5920 & 0.5549 & 0.6956 & 0.6245 \\
\hline
\end{tabular}
\end{table}

\subsection{Ablation Study}

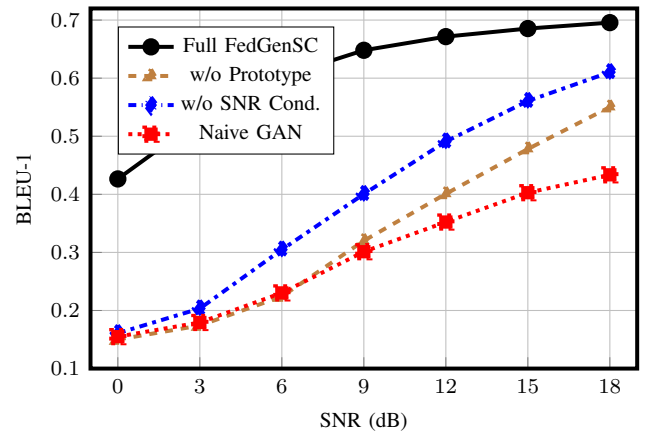
\begin{figure}[t]
\centering
\begin{tikzpicture}
\begin{axis}[
    width=\columnwidth,
    height=0.72\columnwidth,
    xlabel={SNR (dB)},
    ylabel={BLEU-1},
    xlabel style={font=\footnotesize},
    ylabel style={font=\footnotesize},
    tick label style={font=\footnotesize},
    legend style={font=\footnotesize,
        at={(0.05,0.95)}, anchor=north west,
        draw=black, line width=0.4pt},
    xmin=-1, xmax=19,
    ymin=0.10, ymax=0.72,
    xtick={0,3,6,9,12,15,18},
    ytick={0.1,0.2,0.3,0.4,0.5,0.6,0.7},
    grid=both,
    grid style={line width=0.3pt,draw=gray!30},
    major grid style={line width=0.4pt,draw=gray!50},
    tick pos=left, line width=1.5pt]
\addplot[color=black,solid,line width=1.5pt,mark=*,
    mark size=2.5pt,mark options={fill=black,draw=black}]
    coordinates{(0,0.4265)(3,0.5290)(6,0.6028)(9,0.6480)
                (12,0.6715)(15,0.6854)(18,0.6956)};
\addlegendentry{Full FedGenSC}
\addplot[color=brown,dashed,line width=1.5pt,mark=triangle*,
    mark size=2.5pt,mark options={fill=brown,draw=brown}]
    coordinates{(0,0.1500)(3,0.1741)(6,0.2241)(9,0.3201)
                (12,0.4008)(15,0.4780)(18,0.5502)};
\addlegendentry{w/o Prototype}
\addplot[color=blue,dashdotted,line width=1.5pt,mark=diamond*,
    mark size=3pt,mark options={fill=blue,draw=blue}]
    coordinates{(0,0.1609)(3,0.2034)(6,0.3048)(9,0.4005)
                (12,0.4909)(15,0.5607)(18,0.6103)};
\addlegendentry{w/o SNR Cond.}
\addplot[color=red,dotted,line width=1.5pt,mark=square*,
    mark size=2.5pt,mark options={fill=red,draw=red}]
    coordinates{(0,0.1550)(3,0.1794)(6,0.2303)(9,0.3012)
                (12,0.3523)(15,0.4028)(18,0.4336)};
\addlegendentry{Naive GAN}
\end{axis}
\end{tikzpicture}
\caption{Ablation study: BLEU-1 vs.\ SNR (non-IID,
$\alpha{=}0.5$, $K{=}10$ clients, Rayleigh fading). Full
FedGenSC outperforms all ablated variants. Naive GAN federation
performs worst at high SNR, confirming the necessity of
local-only discriminators.}
\label{fig:ablation}
\end{figure}

Table~\ref{tab:ablation_full} and Fig.~\ref{fig:ablation} compare
three ablated variants against full FedGenSC under non-IID
partitioning. Each variant removes exactly one component while
keeping all others intact, isolating the independent contribution
of each design element.

\paragraph{Naive GAN federation}
Aggregating the discriminator causes the most severe degradation:
$-61.8\%$ at 6\,dB, exceeding the no-GAN baseline at 6, 9, 12, and
15\,dB and falling \emph{below} it only at 18\,dB
(Table~\ref{tab:main_results_full}). This is consistent with P1:
incompatible aggregated discriminators can provide inconsistent
adversarial gradients that harm the generator, indicating that
local-only discriminators are necessary in this evaluated setting.

\paragraph{Prototype bank removal}
Removing prototype conditioning causes the largest single-component
drop at 6\,dB, consistent with P2: without a global semantic
reference, each client's generator may specialize locally, and the
aggregated generator can drift toward a compromise that serves no
client well. The decreasing gap at high SNR suggests that
channel-aware correction alone recovers some performance even
without semantic grounding.

\paragraph{SNR conditioning removal}
Removing SNR conditioning produces the smallest drop among the
three ablations, which shrinks further with increasing SNR. At low
SNR, precise channel conditioning is likely important because the
generator must apply large corrections calibrated to the severe
noise level; at high SNR, the smaller drop suggests some channel
adaptation may still occur implicitly through the residual and
adversarial terms alone. This is consistent with P3: SNR
conditioning contributes to accurate channel-adaptive semantic
pre-equalization in the evaluated setting.

\paragraph{Key insights}
All three components contribute independently across the full SNR
range in this evaluated setting: prototype removal and naive GAN
federation cause the largest drops, indicating that semantic
grounding and discriminator locality are both important, while
naive GAN federation additionally falls below the no-GAN baseline
at high SNR, indicating that discriminator aggregation is best
avoided in this setting.

\section{Conclusion}
FedGenSC integrates adversarially trained semantic adaptation into
federated SemCom, addressing three challenges of naive GAN
federation: discriminator aggregation instability, semantic drift,
and channel-agnostic generation. By employing local-only
discriminators, a semantic prototype bank, and SNR conditioning,
the framework improves semantic reconstruction under non-IID data
and heterogeneous channel conditions, achieving up to $58.2\%$
BLEU-1 improvement over the non-IID FedDeepSC baseline on the
Europarl dataset under Rayleigh fading, with SBERT similarity and
ablation results confirming the contribution of each component.
These results highlight the potential of generative,
channel-aware federated SemCom for future wireless systems.


\end{document}